\documentclass[%
 reprint,
 amsmath,amssymb,
 aps,
]{revtex4-2}
 \usepackage{multirow}
\usepackage{booktabs}
\usepackage[version=4]{mhchem}
\usepackage{hyperref}
\usepackage{graphicx}
\usepackage{subcaption} 
\usepackage{dcolumn}
\usepackage{bm}

\begin{document}

\preprint{APS/123-QED}

\title{Effects of geometric configurations in \\ relativistic isobaric collisions at $\sqrt{s_{NN}}=200$ GeV}

\author{Akash Das$^{1}$}
 \email{24pnpo01@iiitdmj.ac.in}
\author{Satya Ranjan Nayak$^{2}$}
 \email{satyanayak@bhu.ac.in}
\author{B. K. Singh$^{1,2}$}%
 \email{bksingh@bhu.ac.in}
 \email{director@iiitdmj.ac.in}
\affiliation{$^{1}$Discipline of Natural Sciences, PDPM Indian Institute of Information Technology Design \& Manufacturing, Jabalpur 482005, INDIA.\\
$^{2}$Department of Physics, Institute of Science,\\ Banaras Hindu University (BHU), Varanasi, 221005, INDIA. }
\date{\today}

\begin{abstract}

In this work, we investigate the role of nuclear deformation ($\beta_{2}$, $\beta_{3}$) and surface diffuseness ($a$) on charged-hadron multiplicity ($N_{\mathrm{ch}}$) and elliptic flow ($v_{2}$) in symmetric isobaric collisions of $^{96}_{44}\mathrm{Ru}+^{96}_{44}\mathrm{Ru}$ and $^{96}_{40}\mathrm{Zr}+^{96}_{40}\mathrm{Zr}$ at $\sqrt{s_{NN}} = 200$~GeV within the HYDJET++ model. Two extreme geometric configurations, tip--tip and body--body, are employed to systematically probe the sensitivity of final-state observables to initial nuclear geometry. A dedicated analysis of pseudorapidity-dependent particle production is carried out to investigate the influence of nuclear deformation on the longitudinal structure of the system. Centrality is determined separately for each configuration to enable a controlled comparison of deformation effects. We find that quadrupole deformation and surface diffuseness produce orientation-dependent modifications of multiplicity and elliptic flow through changes in the effective overlap region. The inclusion of octupole deformation introduces additional asymmetry effects that become most visible in multiplicity distributions, particularly in peripheral body--body and tip--tip Zr+Zr collisions. Variations in $v_{2}$ remain modest and configuration dependent, with the impact of $\beta_{3}$ being more pronounced in body--body collisions. Where available, results are compared with STAR blind-analysis data.

\end{abstract}

\maketitle


\section{\label{sec:level1}INTRODUCTION\protect\\ }

The shape and structure of atomic nuclei play a crucial role in heavy-ion collisions by influencing the initial overlap geometry and the number of participating nucleons. In modeling heavy-ion collisions, each nucleus is described using the Woods--Saxon (WS) density profile, where the surface diffuseness parameter \(a\) controls how gradually the nuclear density falls from its central value to zero at the surface. Nuclear deformation is characterized by parameters such as \(\beta_2\) (quadrupole) and \(\beta_3\) (octupole), which describe deviations from a perfect spherical shape. In particular, Ru exhibits a larger quadrupole deformation (\(\beta_2\)), whereas Zr shows a stronger octupole component (\(\beta_3\)) \cite{Zhang:2021kxj}. In symmetric collisions, these distinct deformations can produce noticeably different initial overlap geometries, resulting in measurable differences in final-state observables like multiplicity and flow.

Charge-Parity (CP) violation is a phenomenon through which we can understand the existence of our matter-dominated universe \cite{Nir:1999mg}. There is evidence of CP violation in weak decay \cite{Fleischer:2024uru,LHCb:2024exp}, but it is still not experimentally proven in strong interactions. Recent experiments in the Relativistic Heavy-Ion Collider (RHIC) and the
Large Hadron Collider (LHC) gives positive hints of possible CP-violating effects in Quark-Gluon Plasma (QGP)
\cite{Collins:1974ky,Shuryak:1978ij}. Originally, the motivation for isobaric collisions (\ce{^{96}_{44}Ru + ^{96}_{44}Ru} and \ce{^{96}_{40}Zr+ ^{96}_{40}Zr}) at RHIC was to explore the possible manifestation of the Chiral Magnetic Effect (CME), which is a signal of local parity (P) and charge-parity (CP) violation in strong interactions~\cite{Feng:2025yte,Fukushima:2008xe,Kharzeev:2004ey,Kharzeev:2007jp,STAR:2023gzg}. In non-central collisions, spectator protons create an intense magnetic field that can interact with chirally imbalanced quarks in the Quark-Gluon Plasma (QGP), inducing charge separation along the field direction. The key idea was that, since isobars have the same mass number, this can lead to a similar background. But due to different proton numbers, it should result in different strengths of the magnetic field and hence different CME signals \cite{Feng:2021oub}.
However, the STAR collaboration’s blind analysis of isobar data \cite{STAR:2021mii} found that the two systems exhibited different background signals due to variations in charged-particle production (\(N_{ch}\)) and flow harmonics (\(v_n\)) \cite{Zhao:2022grq}, primarily originating from differences in their nuclear shapes~\cite{Li:2022bhl}.
Thus, a clear understanding of how deformation affects these observables is essential before isolating any possible CME contributions.

During the fireball’s evolution immediately following a collision, the initial spatial anisotropy translates into final momentum anisotropy~\cite{Kolb:2003dz}.
The anisotropy in particle emission is usually expressed through a Fourier expansion of the azimuthal distribution~\cite{Pandey:2021ofb}:

\begin{equation}
\frac{dN}{d\phi} \propto 1+2
\sum_{n=1}^{\infty} v_n \cos \left[ n
(\phi - \psi_n) \right]                                                  
\end{equation} 

where $\phi$ is the azimuthal angle of the produced particle, $v_n$ is the n'th order fourier coefficient and $\psi_n$ is the angle of the reaction plane.

\begin{table*}[ht]
\centering
\renewcommand{\arraystretch}{1.2}
\begin{tabular}{c|cccc|cccc|cccc}
\hline
\multirow{2}{*}{\textbf{Species}} 
& \multicolumn{4}{c|}{\textbf{Case-1}~\cite{Deng:2016knn}} 
& \multicolumn{4}{c|}{\textbf{Case-2}~\cite{vanderSchee:2023uii}} 
& \multicolumn{4}{c}{\textbf{Case-3}~\cite{Xu:2021vpn}} \\
\cline{2-13}
& $R$ (fm) & $a$ (fm) & $\beta_2$ & $\beta_3$ 
& $R$ (fm) & $a$ (fm) & $\beta_2$ & $\beta_3$ 
& $R$ (fm) & $a$ (fm) & $\beta_2$ & $\beta_3$ \\
\hline
\ce{^{96}_{44}Ru} & 5.085 & 0.46  & 0.158 & 0     & 5.063 & 0.485 & 0.154 & 0     & 5.067 & 0.500 & 0     & 0 \\
\ce{^{96}_{40}Zr} & 5.020 & 0.46  & 0.080 & 0     & 4.960 & 0.536 & 0.062 & 0.202 & 4.965 & 0.556 & 0     & 0 \\
\hline
\end{tabular}
\caption{Woods-Saxon parameters for \ce{^{96}_{44}Ru} and \ce{^{96}_{40}Zr} nuclei.}
\label{tab:WSparameters}
\end{table*}

\begin{figure*}[t]
\centering

\begin{subfigure}[t]{0.48\textwidth}
    \centering
    \includegraphics[width=\linewidth]{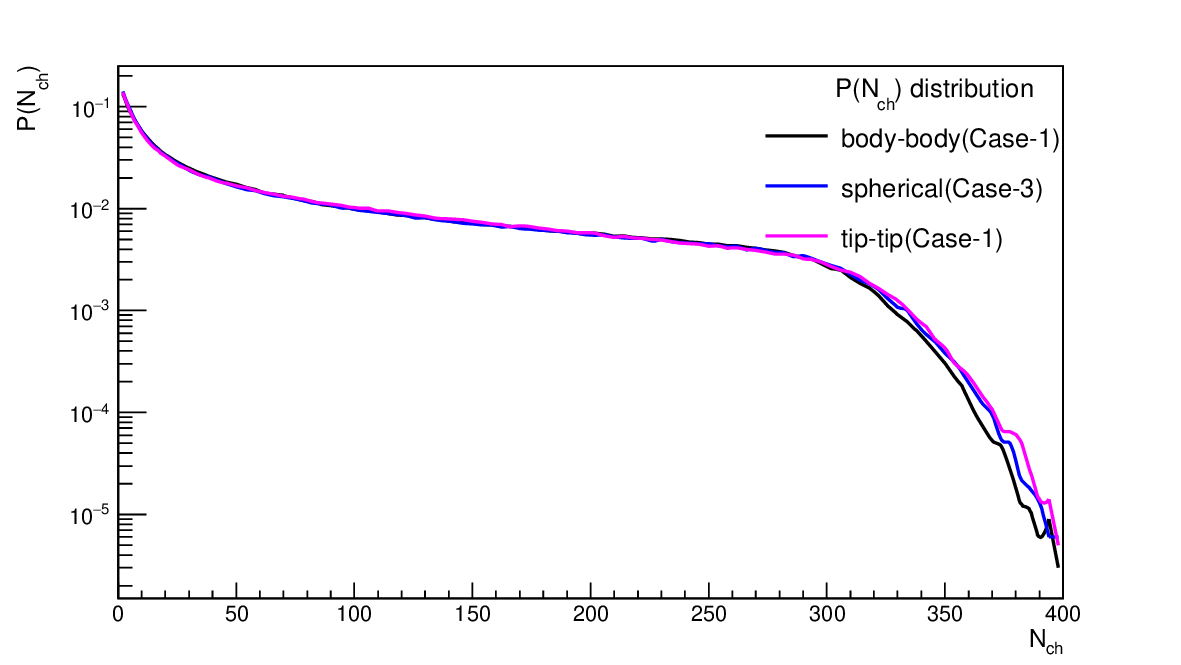}
    \caption{}
    \label{fig:4a}
\end{subfigure}
\hfill
\begin{subfigure}[t]{0.48\textwidth}
    \centering
    \includegraphics[width=\linewidth]{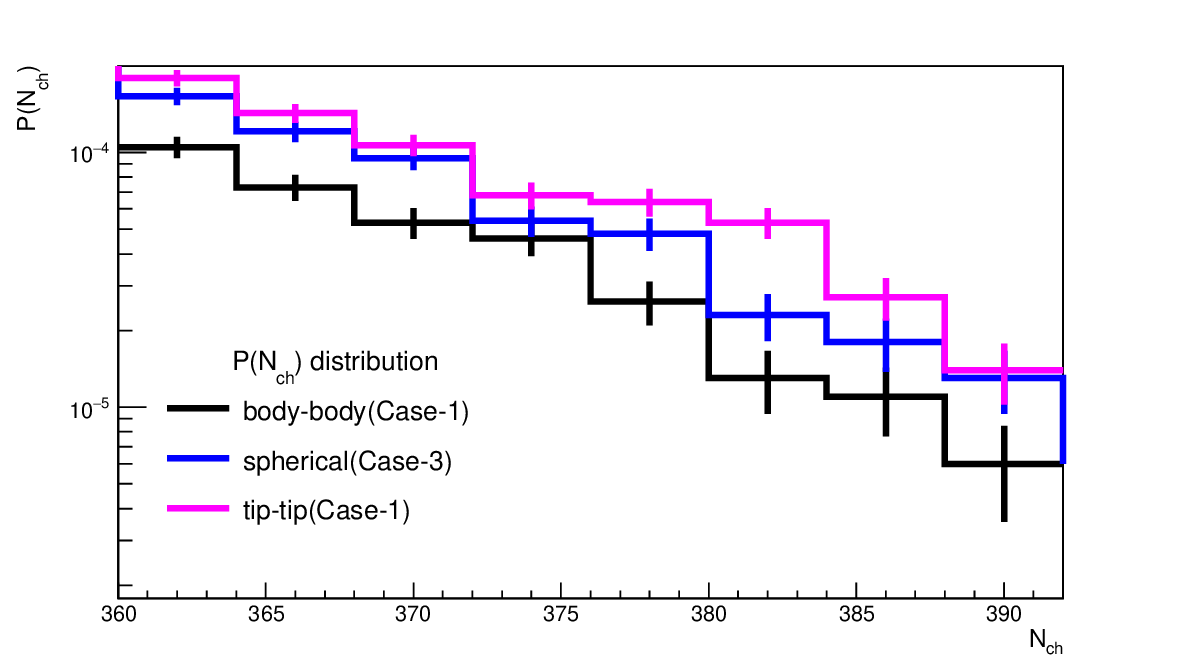}
    \caption{}
    \label{fig:4b}
\end{subfigure}

\caption{Variation of total charged particle multiplicity $(N_\text{ch})$ for body-body (case-1), spherical (case-3) and tip-tip (case-1) collisions.}
\label{fig:your_label}
\end{figure*}

\begin{figure*}[t]
\centering

\begin{subfigure}{0.495\textwidth}
    \includegraphics[width=\linewidth]{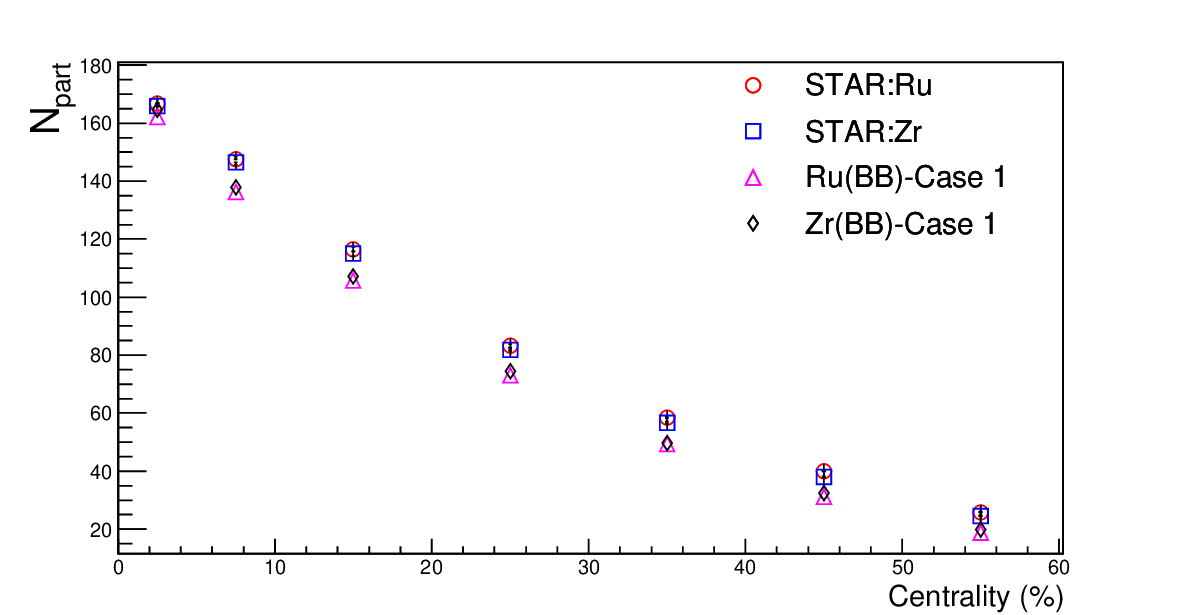}
    \caption{}
    \label{fig:4a}
\end{subfigure}
\hfill
\begin{subfigure}{0.495\textwidth}
    \includegraphics[width=\linewidth]{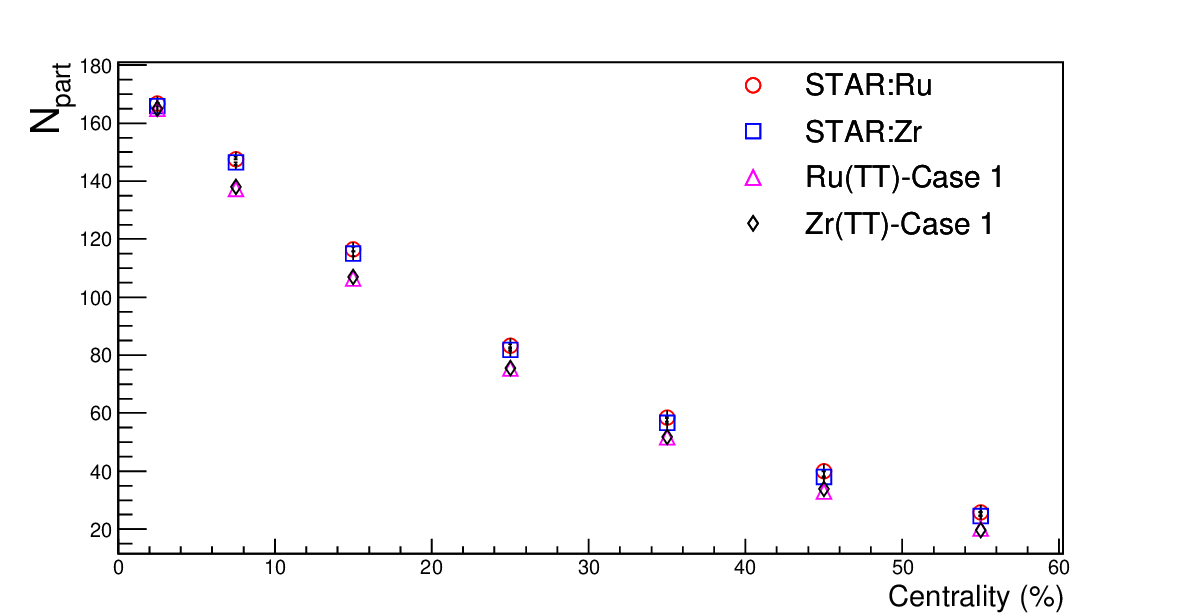}
    \caption{}
    \label{fig:4b}
\end{subfigure}

\vspace{1em}

\begin{subfigure}{0.495\textwidth}
    \includegraphics[width=\linewidth]{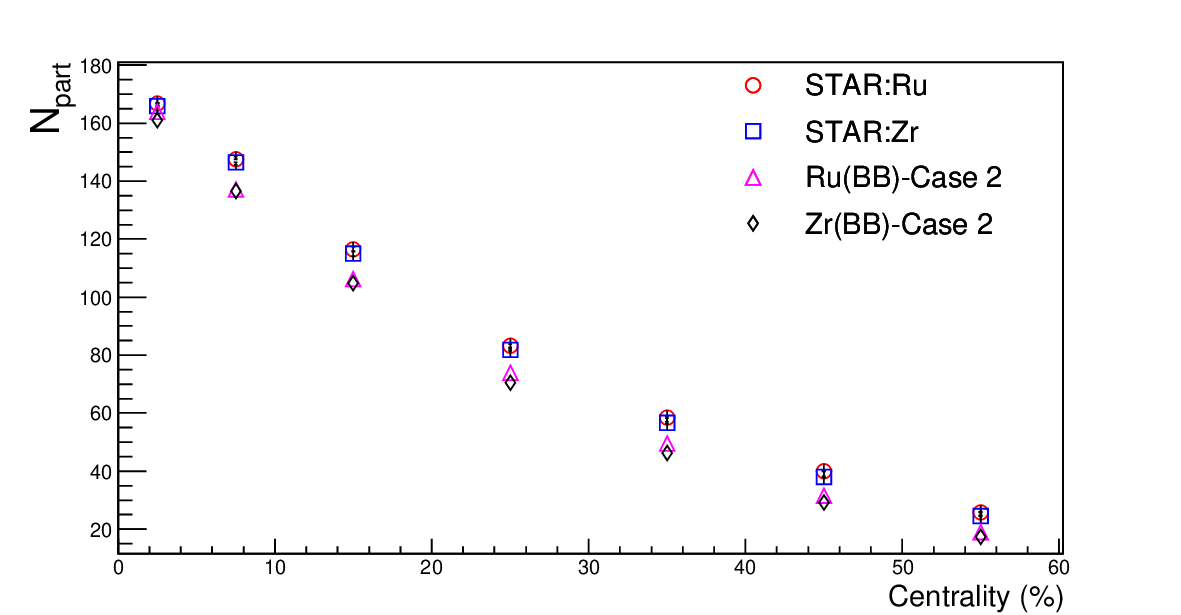}
    \caption{}
    \label{fig:4c}
\end{subfigure}
\hfill
\begin{subfigure}{0.495\textwidth}
    \includegraphics[width=\linewidth]{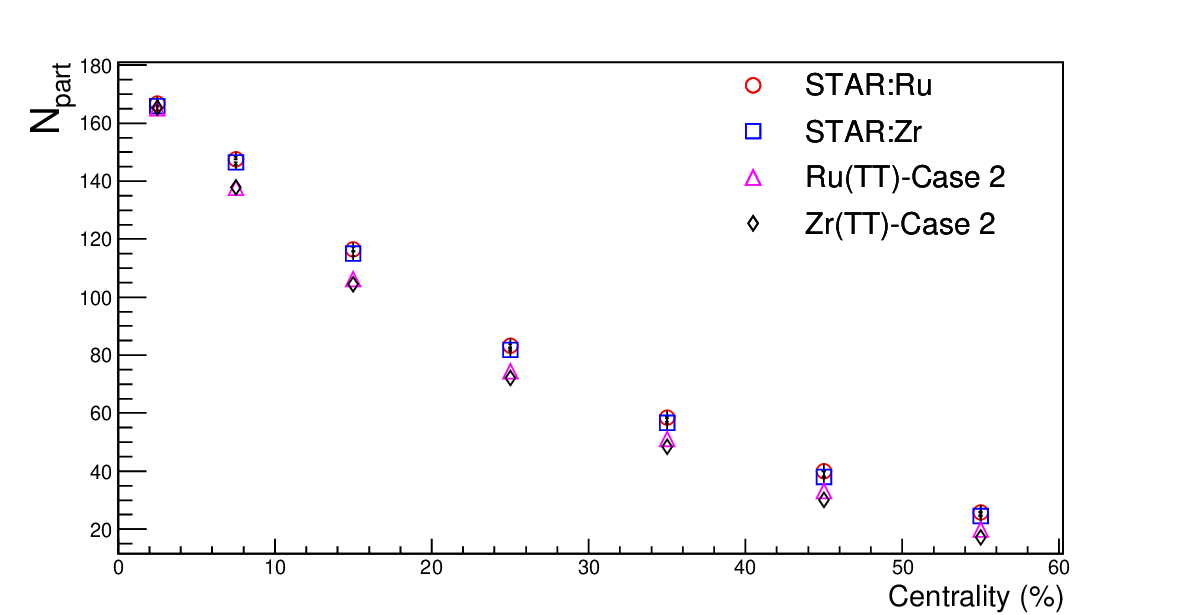}
    \caption{}
    \label{fig:4d}
\end{subfigure}

\vspace{1em}

\begin{subfigure}{0.6\textwidth}
    \centering
    \includegraphics[width=\linewidth]{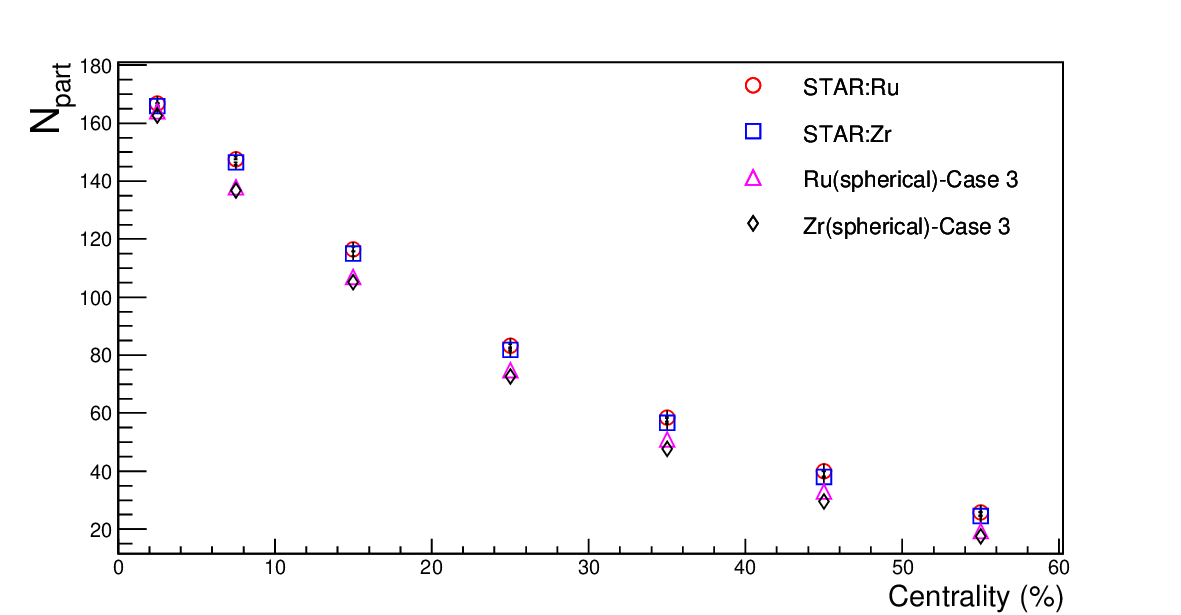}
    \caption{}
    \label{fig:4e}
\end{subfigure}

\caption{Variation of $N_\text{part}$ with respect to centrality for (a) body-body (case-1), (b) tip-tip (case-1), (c) body-body (case-2), (d) tip-tip (case-2), (e) spherical (case-3) collisions.}
\label{fig:fig4}
\end{figure*}

We begin our analysis with a spherical nuclear configuration (i.e., no deformation) and subsequently include deformation by considering two extreme geometrical configurations of the colliding nuclei: tip-tip and body-body. Although intermediate orientations can be included in the model, as we discuss them in the model description, the tip-tip and body-body cases represent limiting configurations that most clearly amplify or suppress deformation-induced effects.
The objective is to investigate the effects of these specific geometrical configurations by analyzing the charged hadron pseudorapidity density ($\frac{dN_{\text{ch}}}{d\eta}$), the number of charged hadrons ($N_{\text{ch}}$), and the elliptic flow coefficients ($v_2$), with particular attention to configuration-dependent yield differences and longitudinal stability. Centrality classes are determined independently for each nuclear orientation in order to account for orientation-dependent overlap geometries.

This article is organized as follows: Section~II describes the HYDJET++ model. Section~III presents the results and discussion under three parts: (A) pseudorapidity distribution, (B) \(N_{ch}\) distribution analysis, and (C) \(v_2\) ratio. Section~IV concludes with a summary of our findings.

\begin{figure*}[t]
\centering

\begin{subfigure}{0.495\textwidth}
    \includegraphics[width=\linewidth]{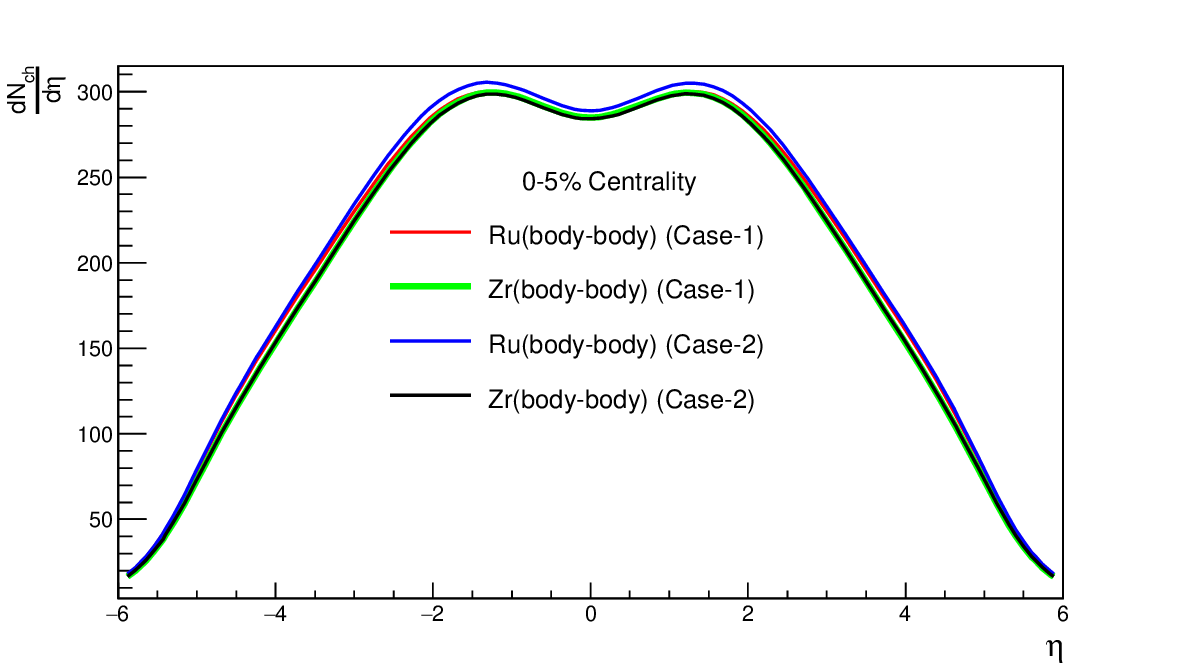}
    \caption{}
    \label{fig:8a}
\end{subfigure}
\hfill
\begin{subfigure}{0.495\textwidth}
    \includegraphics[width=\linewidth]{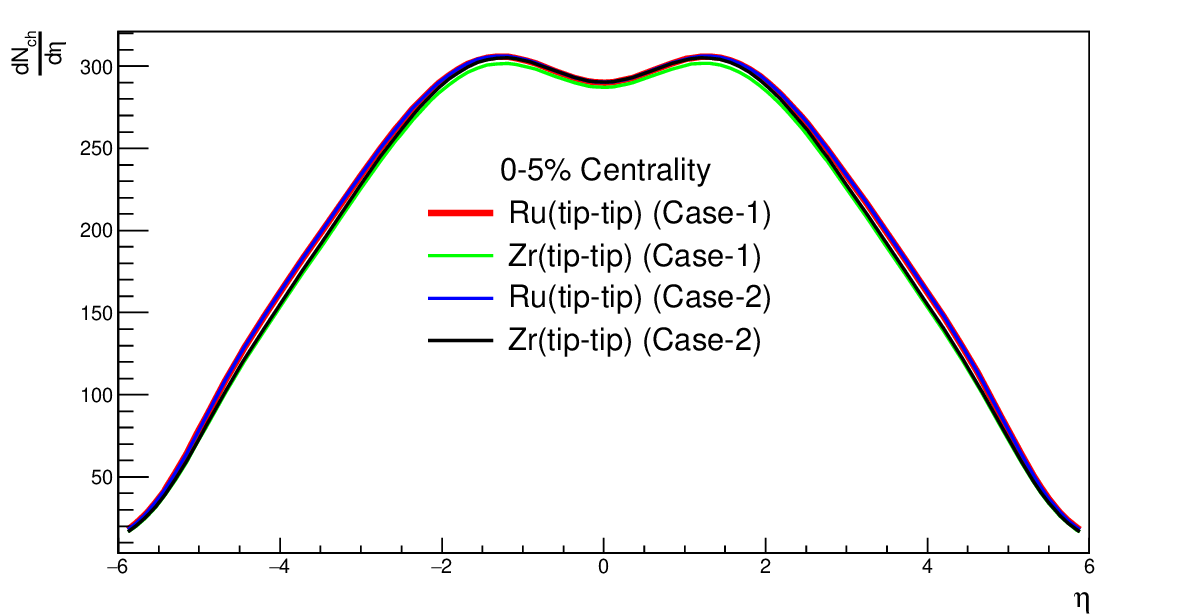}
    \caption{}
    \label{fig:8b}
\end{subfigure}

\vspace{1em}

\begin{subfigure}{0.495\textwidth}
    \includegraphics[width=\linewidth]{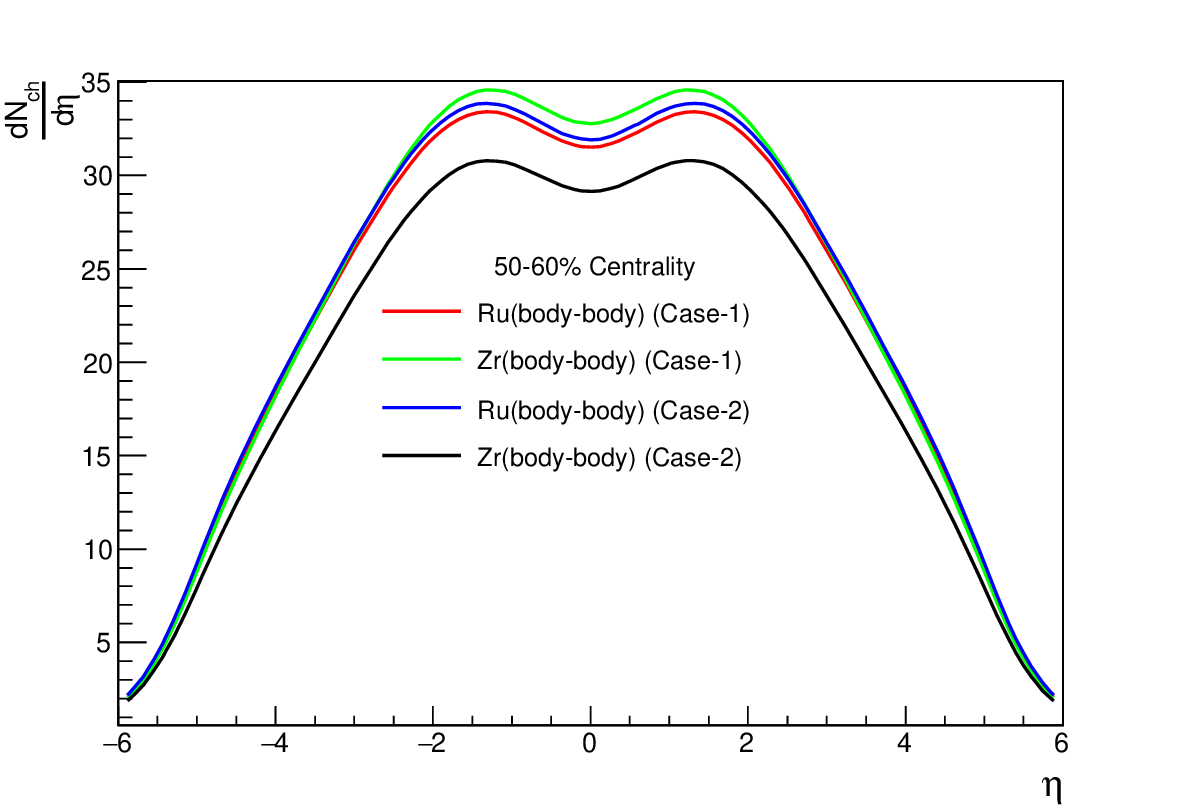}
    \caption{}
    \label{fig:8c}
\end{subfigure}
\hfill
\begin{subfigure}{0.495\textwidth}
    \includegraphics[width=\linewidth]{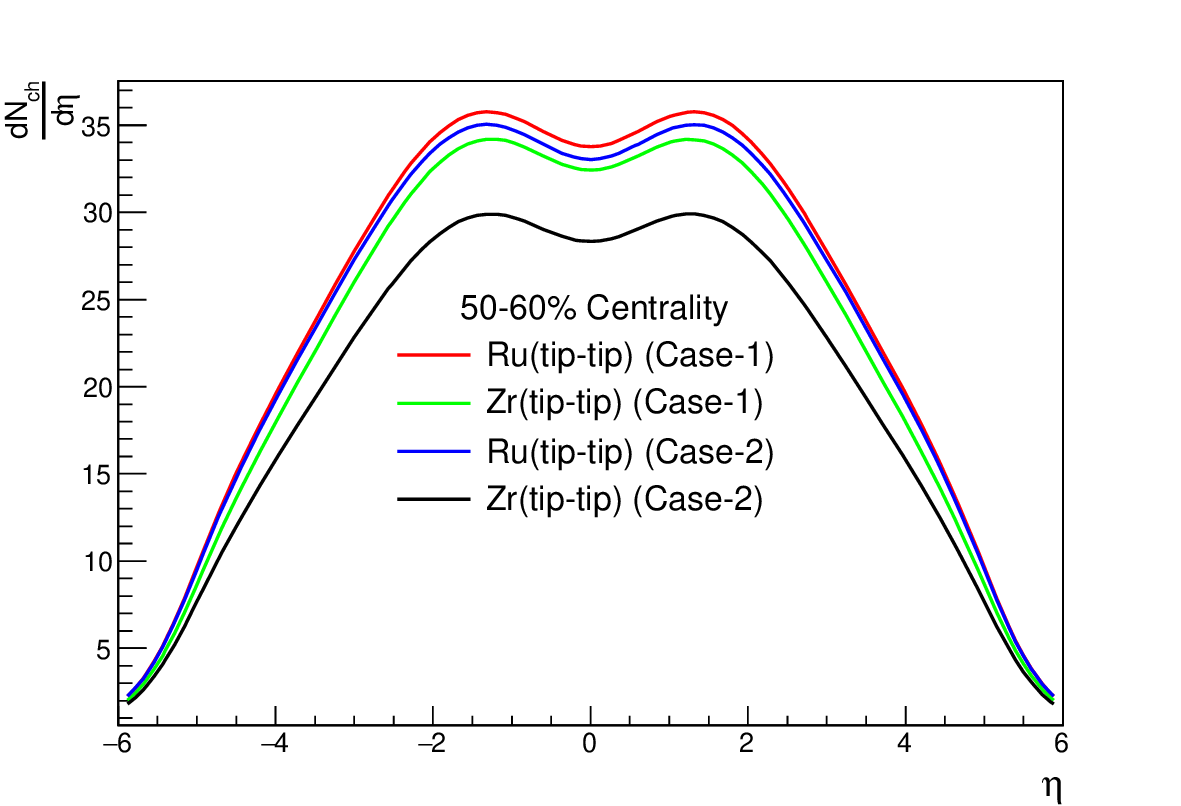}
    \caption{}
    \label{fig:8d}
\end{subfigure}

\caption{Variation of $\frac{dN_\text{ch}}{d\eta}$ with respect to $\eta$ for (a) body-body ($0$-$5$\% centrality), (b) tip-tip ($0$-$5$\% centrality), (c) body-body ($50$-$60$\% centrality), (d) tip-tip ($50$-$60$\% centrality) collisions.}
\label{fig:fig8_first}
\end{figure*}

\vspace{2em}

\begin{figure*}[t]
\centering

\begin{subfigure}{0.495\textwidth}
    \includegraphics[width=\linewidth]{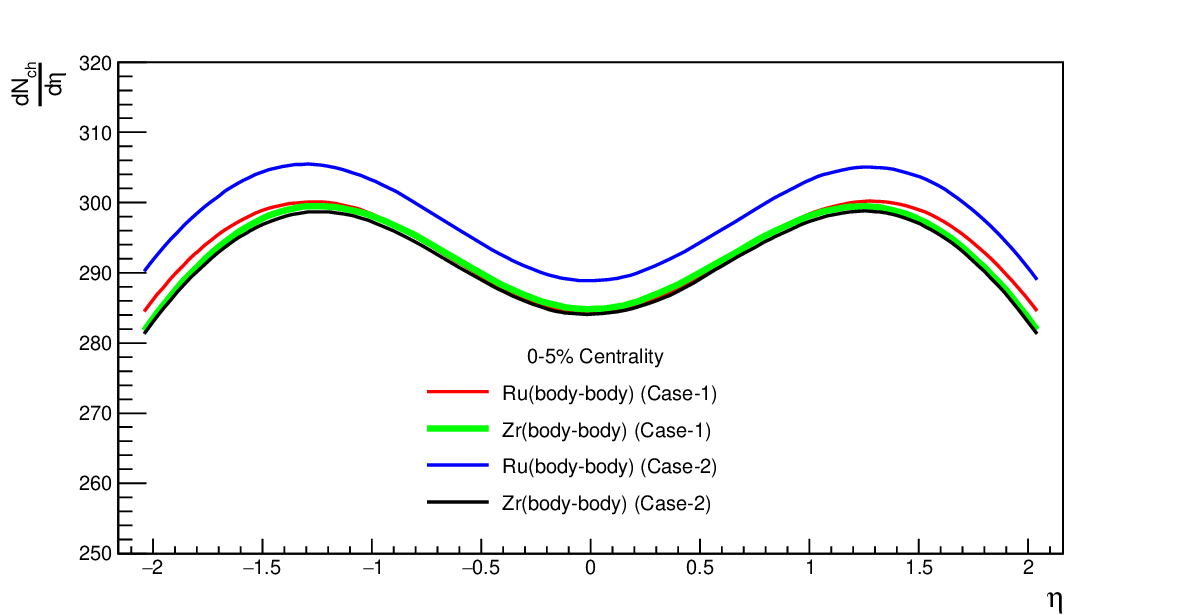}
    \caption{}
    \label{fig:8e}
\end{subfigure}
\hfill
\begin{subfigure}{0.495\textwidth}
    \includegraphics[width=\linewidth]{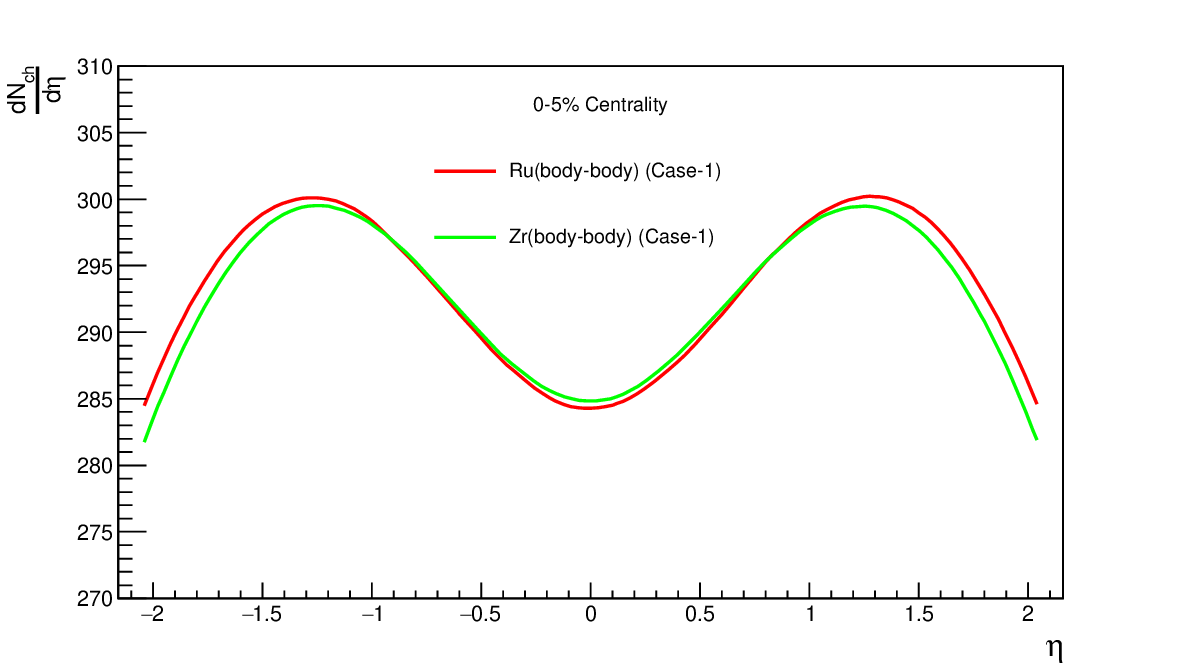}
    \caption{}
    \label{fig:8f}
\end{subfigure}

\vspace{1em}

\begin{subfigure}{0.495\textwidth}
    \includegraphics[width=\linewidth]{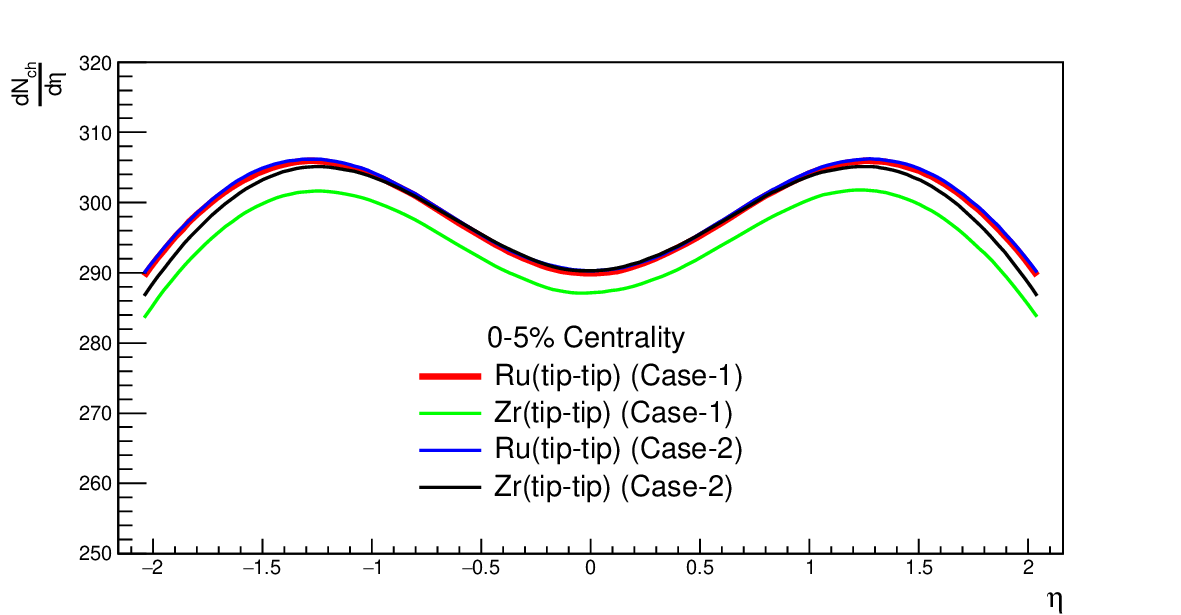}
    \caption{}
    \label{fig:8g}
\end{subfigure}
\hfill
\begin{subfigure}{0.495\textwidth}
    \includegraphics[width=\linewidth]{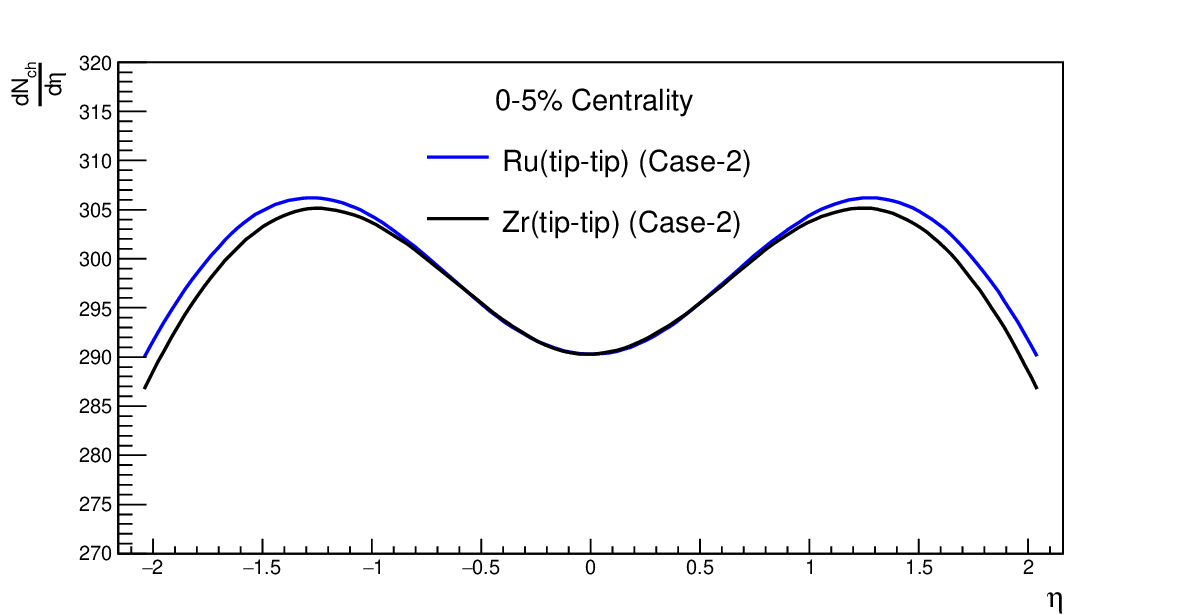}
    \caption{}
    \label{fig:8h}
\end{subfigure}

\caption{A magnified view of Fig. 3(a) is shown in (a) and (b) here. Again, an enlarged view of Fig. 3(b) is presented in (c) and (d).}
\label{fig:fig8_second}
\end{figure*}

\begin{figure*}[t]
\centering
\begin{subfigure}{0.495\textwidth}
    \includegraphics[width=\linewidth]{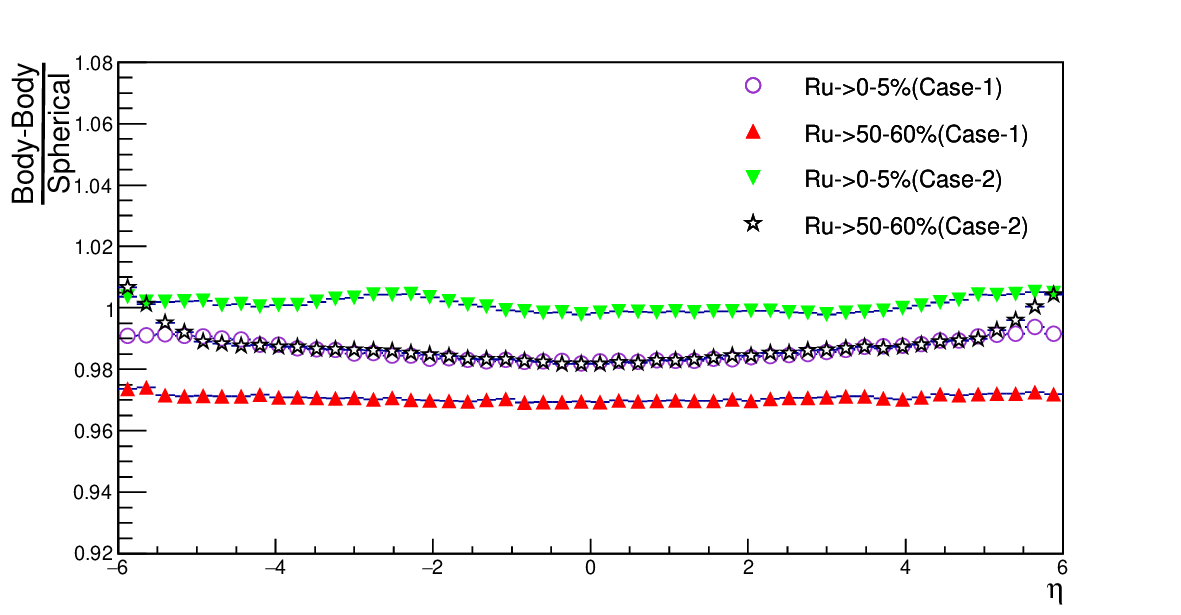}
    \caption{}
    \label{fig:4a}
\end{subfigure}
\hfill
\begin{subfigure}{0.495\textwidth}
    \includegraphics[width=\linewidth]{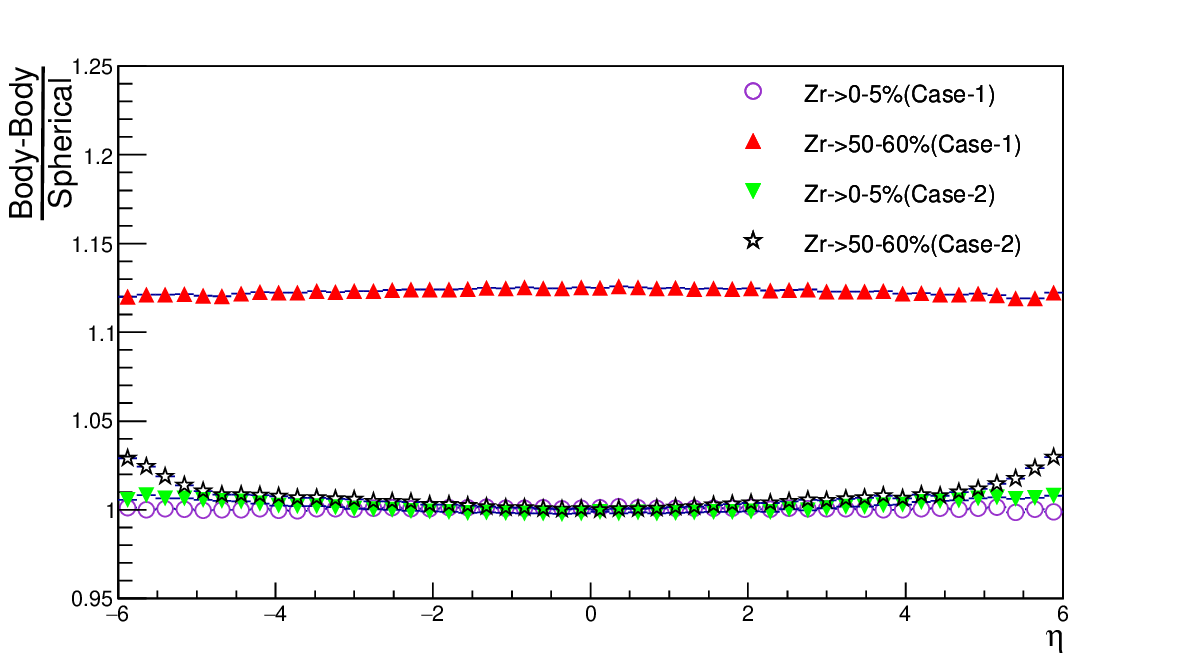}
    \caption{}
    \label{fig:4b}
\end{subfigure}

\vspace{1em}

\begin{subfigure}{0.495\textwidth}
    \includegraphics[width=\linewidth]{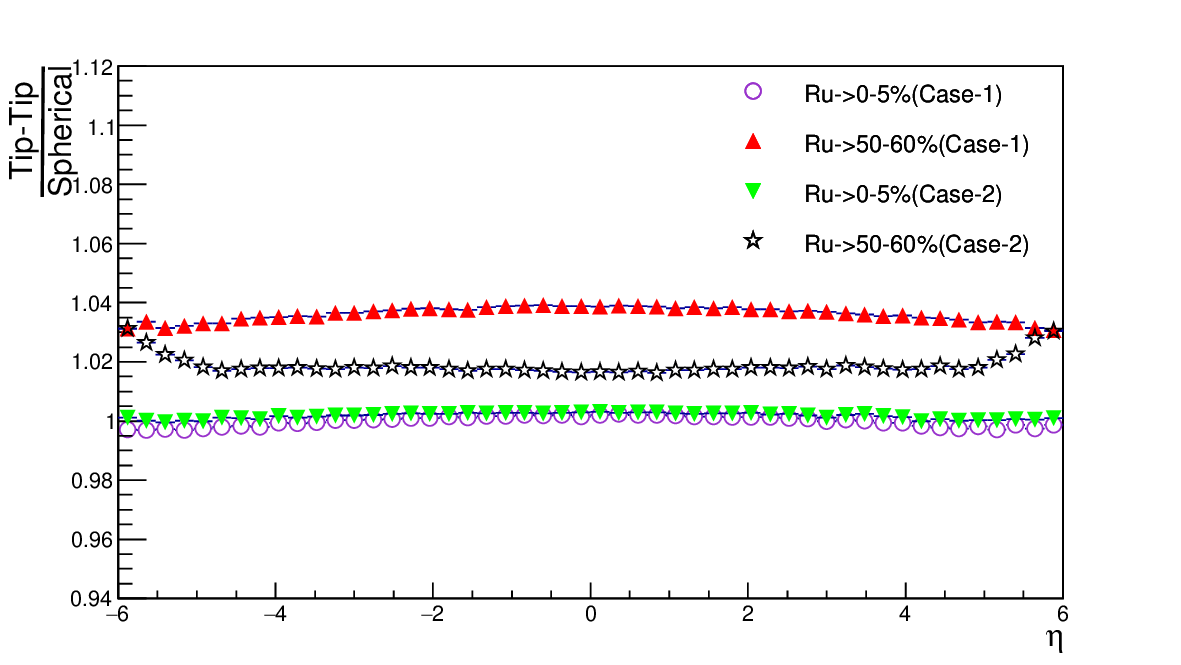}
    \caption{}
    \label{fig:4c}
\end{subfigure}
\hfill
\begin{subfigure}{0.495\textwidth}
    \includegraphics[width=\linewidth]{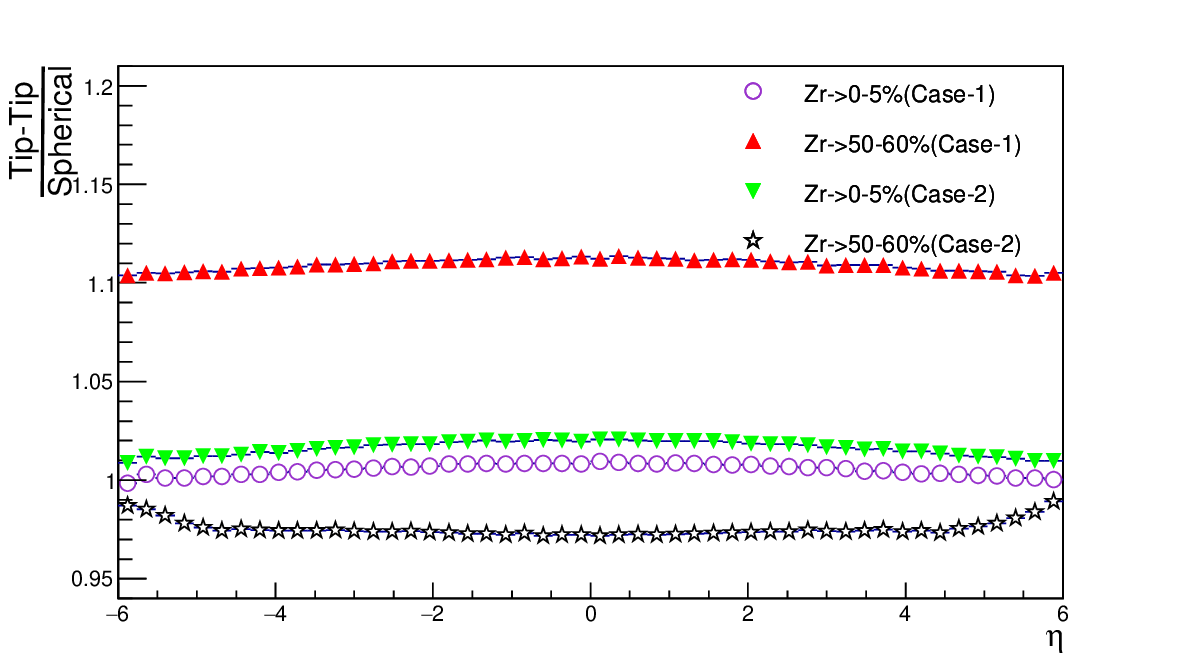}
    \caption{}
    \label{fig:4d}
\end{subfigure}

\caption{Ratio of pseudorapidity distributions for deformed to spherical configurations: (a) body-body to spherical for Ru, (b) body-body to spherical for Zr, (c) tip-tip to spherical for Ru, and (d) tip-tip to spherical for Zr.}
\label{fig:fig4}
\end{figure*}

\section{HYDJET++}

The HYDJET++ is a Monte Carlo event generator, commonly used for the simulation of relativistic heavy-ion collisions. The HYDJET++ model superposes both the soft hydro-type state and the hard state independently, yet simultaneously. The model uses  Bjorken's solution of ideal hydrodynamics \cite{Bjorken:1982qr} for space-time evolution of the fireball, hence the model calculations better describe mid-rapidity data. At forward rapidities, one should use Landau hydrodynamics \cite{Steinberg:2004vy} for a more accurate description of medium expansion. A detailed physics framework of the model and simulation can be seen in these articles \cite{Lokhtin:2008xi, Bravina:2017rkp}. The key points are outlined below:-

\textbf{Hard Part}:- HYDJET++ models the hard interactions using the PYTHIA Quenched (PYQUEN) partonic energy loss model \cite{Lokhtin:2005px}. The initial particle spectra are generated by PYTHIA 6 \cite{Sjostrand:2006za}. Later, PYQUEN modifies the initial particle spectra by generating binary nucleon-nucleon collisions at a specified impact parameter, following the Glauber Model \cite{Miller:2007ri}. In PYTHIA, the parameter $p_T^{\text{min}}$ separates the hard part from the soft part. If the momentum transfer in individual interactions is greater than $p_T^{\text{min}}$, i.e, $p_T$$>$$p_T^{\text{min}}$, then these interactions are considered as hard processes by PYQUEN. Moreover for the interactions $p_T$$<$$p_T^{\text{min}}$, they are considered as soft interactions. PYQUEN does an event-by-event calculation on re-scattering of partons via integrating the radiation and collisional energy losses \cite{Bjorken:1982qr, Baier:2001qw} inside the medium. Finally, hadronization is done according to the Lund string fragmentation model \cite{Andersson:1997xwk}.

\textbf{Soft Part}:- The soft part of HYDJET++ is a thermal hadronic state, which is generated on the chemical and thermal freeze-out hypersurfaces obtained from a parametrization of relativistic hydrodynamics with preset freeze-out conditions \cite{Amelin:2006qe, Amelin:2007ic}. The QGP created in collision reaches a local equilibrium just after a short period of time $(<1  fm/c)$ and further expands hydrodynamically. In HYDJET++, chemical and thermal freezeouts are assumed to happen separately because at chemical freezeout, the density of the particles is too high to consider them as free streaming. So, a more complex but relevant condition of $T_{ch} \leq T_{th}$ is used. The system expands hydrodynamically with frozen chemical composition in between the two freezeout conditions, then it cools down, and hadrons start moving freely after they reach the thermal freezeout temperature.

\textbf{Elliptic Flow}:- Shortly after the collision (preferably peripheral), an asymmetric almond shape is formed in the overlap region. It produces a pressure gradient, generally stronger towards the shorter axis of the ellipse. This initial anisotropy transfers into momentum anisotropy, and the flow phenomenon is called elliptic flow \cite{Sorge:1996pc}. And $v_2$ can be defined in terms of particle momenta:

\begin{equation}
v_2 = \left\langle \frac{p_x^2 - p_y^2}{p_x^2 + p_y^2} \right\rangle 
= \left\langle \frac{p_x^2 - p_y^2}{p_T^2} \right\rangle
\end{equation}

Unlike microscopic transport models such as AMPT~\cite{Lin:2004en} or full hydrodynamic models like MUSIC~\cite{Schenke:2010nt}, which explicitly simulate the dynamical evolution of the medium, \textsc{HYDJET++} employs a phenomenological parameterization of the freeze-out hypersurface to reproduce collective flow patterns. In this framework, the elliptic flow arises from two coefficients:- the spatial anisotropy ${\epsilon_2}$ and the momentum anisotropy ${\delta_2}$. These coefficients describe, respectively, the elliptic modulation of the final freeze-out hypersurface and the alteration of the flow rapidity profile. In the default implementation, ${\epsilon_2}$ and ${\delta_2}$ are treated as adjustable parameters in HYDJET++. However, in the present study,  they are defined as functions of the impact parameter and nuclear deformation parameters ($\beta_2$, $\beta_3$), ensuring that the freeze-out geometry reflects the initial nuclear shape and collision orientation. Here, the two parameters are treated as done in the corresponding paper \cite{Pandey:2021ofb}.

\textbf{Deformation}:- As we have mentioned previously, Ru has intrinsic quadrupole deformation and Zr has octupole \cite{Zhang:2021kxj}. But the default HYDJET++ model doesn't have a deformed nuclear density profile. So, the deformed nuclear density function was included. The Modified Woods Saxon distribution function in spherical polar coordinates can be expressed as:

\begin{equation}
\begin{aligned}
\rho(r,\theta,\psi) &= \frac{\rho_0}{1 + \exp\left(\frac{r - R(1 + \beta_2 Y_{20} + \beta_3 Y_{30})}{a}\right)}
\end{aligned}
\end{equation}

\vspace{-1em}

where, $\rho_0 = \rho_0^{\text{const}} + \textit{correction}$, $\rho_0^{\text{const}} = \frac{\text{Mass}}{\text{Volume}} = \frac{3 A}{4\pi R_A^3} = \frac{3}{4\pi R_l^3}$, ${R_A} = R(1 + \beta_2 Y_{20} + \beta_3 Y_{30})$, ${R_l} = {R_0}(1 + \beta_2 Y_{20} + \beta_3 Y_{30})$, $R = R_0 A^{1/3}$ and $R_0 = 1.15fm$. The correction term is calculated as $\rho_0 (\pi f / R_A)^2$
, where $Y_{20} = \sqrt{\frac{5}{16\pi}} \left(3\cos^2\theta - 1\right)$, and $Y_{30} = \sqrt\frac{7}{16 {\pi}}(5\cos^3\theta - 3\cos\theta )$ are the spherical harmonics. 

The HYDJET++ model uses a cylindrical polar coordinate system. So we transformed this nuclear density profile function from spherical $(r,\theta,\psi)$ to cylindrical coordinates $(\rho,\psi,z)$. The relationships of $\theta = \tan^{-1} (z/r)$ and $\theta = \tan^{-1} (r/z)$ for body-body and tip-tip configurations, respectively, are used for the coordinate transformation. Here, \( r \) is used instead of \( \rho \) in the cylindrical polar coordinate system to avoid possible confusion with the nuclear density function $(\rho)$. By changing $\psi$, one can get many other configurations. However, in this study, we are limiting ourselves to body-body and tip-tip configurations only. 

\begin{figure}[htbp]
\centering

\begin{minipage}[t]{0.48\textwidth}
    \centering
    \includegraphics[width=\linewidth]{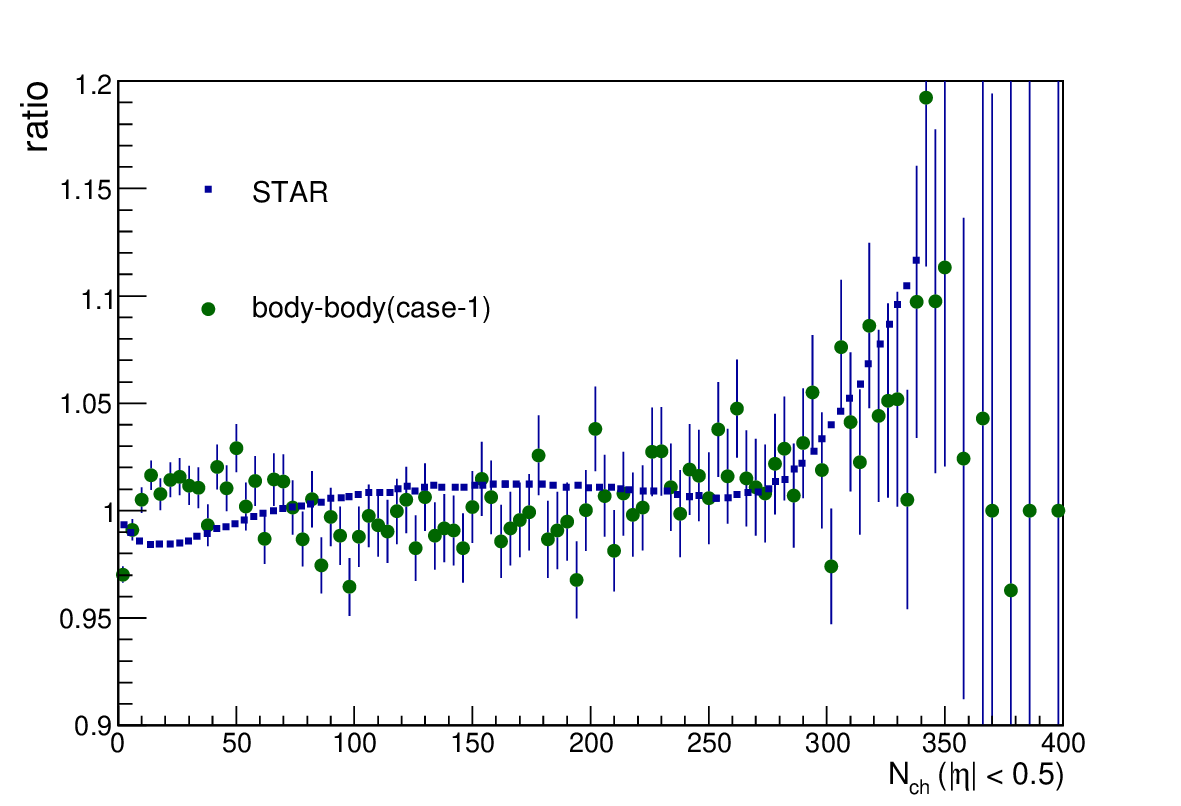}
    \caption*{(a)}

    \vspace{0.2cm}
    \includegraphics[width=\linewidth]{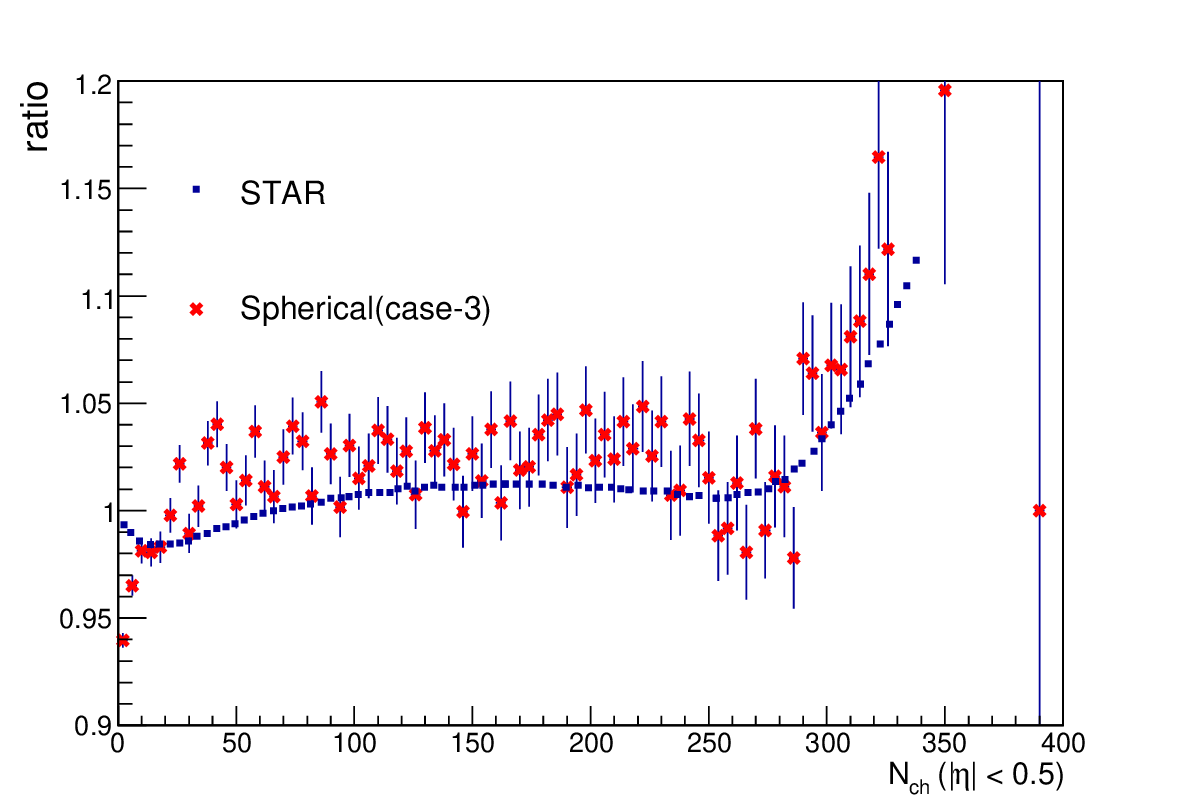}
    \caption*{(b)}

    \vspace{0.2cm}
    \includegraphics[width=\linewidth]{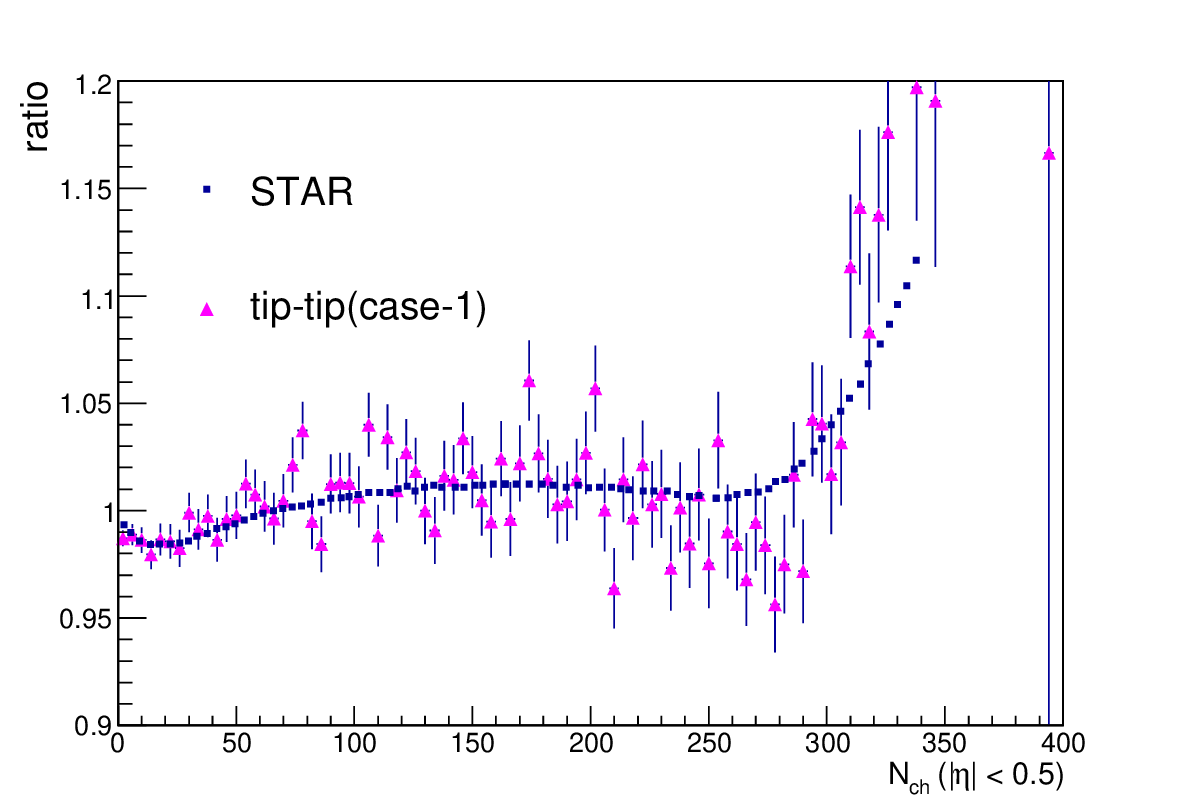}
    \caption*{(c)}
\end{minipage}

\caption{$\frac{P(N_{\mathrm{ch}})^{\text{Ru}}}{P(N_{\mathrm{ch}})^{\text{Zr}}}$ distributions for (a) body-body (case-1), (b) spherical (case-3), and (c) tip-tip (case-1) configurations.}
\label{fig:fig_column1}
\end{figure}

\begin{figure}[htbp]
\centering

\begin{minipage}[t]{0.48\textwidth}
    \centering
    \includegraphics[width=\linewidth]{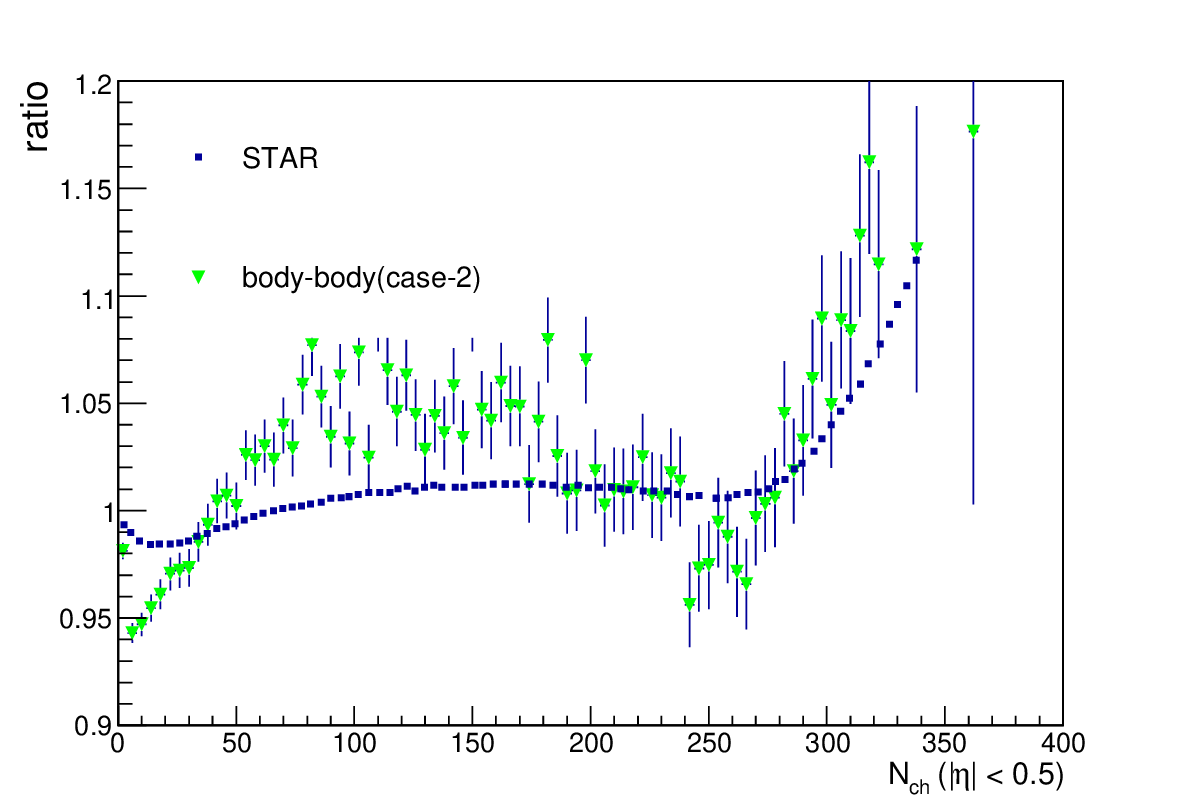}
    \caption*{(a)}

    \vspace{0.2cm}
    \includegraphics[width=\linewidth]{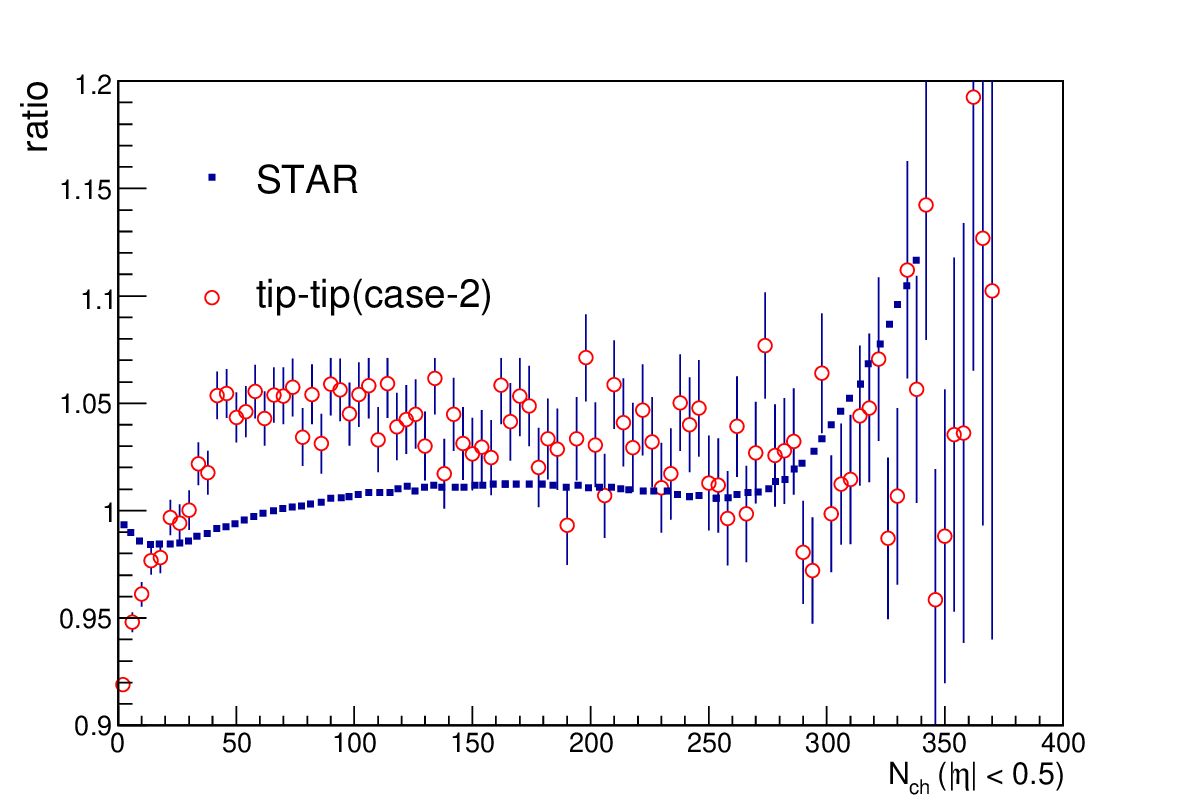}
    \caption*{(b)}
\end{minipage}

\caption{$\frac{P(N_{\mathrm{ch}})^{\text{Ru}}}{P(N_{\mathrm{ch}})^{\text{Zr}}}$ distributions for (a) body-body (case-2) and (b) tip-tip (case-2).}
\label{fig:fig_column2}
\end{figure}

\section{Results and discussions}

The extreme configurations, such as tip–tip and body–body in isobar collisions, provide insight into the influence of nuclear structure on the dynamics of heavy-ion collisions. These extreme configurations can enhance or suppress geometry-driven observables associated with particle production and collective flow. A higher octupole deformation causes a reflection asymmetry in the shape of Zr \cite{Zhang:2021kxj}. Given that the octupole deformation study in the low-energy zone is associated with significant uncertainties \cite{Jia:2022ozr}, we can employ high-energy collisions to probe nucleon configurations event-by-event and get insight into nuclear deformations. 

The introduction of deformation parameters modifies the nuclear density distribution and, consequently, the effective nuclear radius. In the spherical case, the radius is given by a simple empirical formula \( R_0 = r_0 A^{1/3} \), but in the deformed case, it is dependent on \( \theta \). The spherical harmonics are used to describe the nuclear shape. The detailed mathematical formulation can be found in the corresponding papers  \cite{Horiuchi:2021cas, Geldhof2022}. To enable a meaningful comparison with experimental results, we consider a spherical configuration (case-3 in Table-$\mathrm{I}$) based on parameters from density functional theory calculations~\cite{Xu:2021vpn}. In addition, two deformed configurations are included:- case-1 from~\cite{Deng:2016knn} and case-2 from~\cite{vanderSchee:2023uii}, as also summarized in Table-$\mathrm{I}$.

In Fig. 1(a), the $N_{\text{ch}}$ distribution of Ru for body-body, spherical, and tip-tip configurations is shown. For better resolution, we have shown the tail portion separately in Fig. 1(b). The tail regions of the charged hadron multiplicity distributions exhibit noticeable differences across configurations. It means that when the nuclear shape differs, or nuclei collide in different specific orientations, then the same centrality class corresponds to different impact parameters, and hence different numbers of participants ($N_{\text{part}}$) are being produced. The $N_{\text{part}}$ distribution of all those is displayed in Fig. 2. In HYDJET++, the impact parameter range is provided as an input to define the desired centrality class for the generated events. Since the multiplicity distributions differ across configurations, centrality selection was performed independently for each configuration. The centralities were assigned based on the multiplicity of charged hadrons using the integral fraction method \cite {Miller:2007ri}, and the impact parameters corresponding to the $<N_{ch}>$ values were used as inputs.

\subsection{Pseudorapidity distribution}

The pseudorapidity distribution is the measure of the charged particle yield per unit pseudorapidity interval. It gives information about the fireball's evolution. A broader pseudorapidity distribution is generally associated with larger particle production over an extended longitudinal range, while a narrower distribution reflects a more limited particle yield. The pseudorapidity distribution typically depends on the collision centrality and shape of the colliding nuclei. 

In Fig. 3, we present the pseudorapidity distributions of charged hadrons produced in isobaric collisions of Ru+Ru and Zr+Zr for the most central ($0$–$5$\%) and peripheral ($50$-$60$\%) collisions. Ru has slight differences in the values of $a$ and $\beta_2$ between case-1 and case-2. In contrast, Zr exhibits larger variations in both $a$ (from $0.46$ to $0.536$~fm) and $\beta_2$ (from $0.08$ to $0.062$) in the two cases, along with the inclusion of $\beta_3 = 0.202$ in case-2. The deformation parameters ($\beta$) directly influence the nuclear geometry, while $a$ controls the thickness of the overlapping surface. A larger $a$ makes the edge of the nucleus more diffuse, thereby modifying the overlap region during collisions. To improve visualization, the plots of Figs. 3(a) and 3(b) are enlarged and displayed separately in Fig. 4. 

Figures 3(a) and 3(b) depict the pseudorapidity distribution $\frac{dN_{\text{ch}}}{d\eta}$ as a function of ${\eta}$ for the $0$-$5$\% centrality range. At ${\eta}=0$, the central ($0$–$5$\%) body–body collisions with the case-1 parameter set yield a higher charged-hadron multiplicity for Zr than for Ru. Since $a$ is identical for both nuclei in this case, the main difference arises from $\beta_2$. Zr, having a smaller $\beta_2$, is closer to spherical, resulting in a larger central overlap zone and, therefore, a higher multiplicity. In contrast, Fig.~4(a) demonstrates that, within the same centrality range, body–body collisions of Ru with the case-2 parameters yield a higher overall charged-particle multiplicity across the pseudorapidity range. This behavior can be attributed to Ru’s denser core, resulting from its smaller value of $a$.

The $\frac{dN_{\text{ch}}}{d\eta}$ distributions for tip-tip collisions in the most central class ($0$–$5$\%) exhibit a noticeable overlap between Ru and Zr around $\eta = 0$ in case-2. Ru has a comparatively larger quadrupole deformation parameter ($\beta_2=0.154$) than that of Zr ($\beta_2=0.062$). A larger value of $\beta_2$ in tip-tip collisions results in a more elongated nuclear shape along the beam axis, which enhances the production of charged particles ($N_{\mathrm{ch}}$). A higher surface diffuseness ($a$) also contribute to the increased effective overlap volume, particularly in the central collision zone (Zr: $a=0.536$ and Ru: $a=0.485$). While Ru benefits from a higher value of $\beta_2$, Zr compensates for it with a larger value of $a$ and asymmetry in shape ($\beta_3$) that is present in Zr. Thus, the observed overlap in the $\frac{dN_{\text{ch}}}{d\eta}$ distributions suggests that the presence of $\beta_3$ in Zr may partially compensate for its smaller quadrupole deformation in charged-hadron production. In contrast, while considering case-1, Ru's more prominent elliptic form causes it to produce more particles than Zr in central ($0$–$5$\%) tip-tip collisions [Fig. 4(c)] over the whole pseudorapidity range.

In peripheral collisions ($50$–$60$\% centrality) for the body–body case [Fig. 3(c)], Ru exhibits a higher $N_{\mathrm{ch}}$ value across the full $\eta$ range, when case-2 parameter sets are used for both nuclei. Since Zr has more surface diffuseness ($a$) and less elongation ($\beta_2$), Zr should have a higher value of $N_{\mathrm{ch}}$. A possible contributing factor is the asymmetry from $\beta_3$ in case-2, which may reduce the symmetry of the overlap by introducing surface bulges. This in turn decreases the charged hadron production in peripheral body–body collisions of Zr. This trend changes when the case-1 parameter set is employed. In body--body collisions of case--1, Zr exhibits a slightly higher value of $N_{\mathrm{ch}}$ around $\eta = 0$. Since the surface diffuseness parameter $a$ is identical for both Ru and Zr in this case, the observed difference can be attributed primarily to residual geometric effects arising from their different deformation parameters. Ru has a larger quadrupole deformation parameter ($\beta_{2} = 0.158$) compared to Zr ($\beta_{2} = 0.08$). However, in peripheral body--body collisions, the stronger elongation of Ru leads to a reduction in the effective transverse overlap region. In contrast, Zr, with its more compact shape and the same surface diffuseness, provides a slightly larger effective overlap, resulting in a marginally higher charged-particle multiplicity.

Now, analyzing Fig. 3(d), for peripheral ($50$-$60$\%) tip-tip collisions in case--1, Ru exhibits a slightly higher $N_{\mathrm{ch}}$ than Zr. Since the surface diffuseness parameter ($a$) is identical for both nuclei in this case, the observed difference can be attributed primarily to the larger quadrupole deformation ($\beta_2$) of Ru. This suggests that quadrupole deformation plays a key role in determining the effective overlap and particle production in peripheral tip-tip collisions for this configuration. In contrast, for case-2, the Zr (tip-tip) curve lies below all other cases in the $50$-$60$\% centrality bin. This behavior is non-trivial, as Zr has a larger value of $a$, which would be expected to increase the overlap zone, and hence increase $N_{\mathrm{ch}}$. The observed suppression suggests that the octupole deformation ($\beta_3$) introduces shape asymmetry that reduces the effective overlap in this configuration. Therefore, in peripheral tip--tip collisions for case-2, the enhancement due to surface diffuseness is partially offset by the asymmetry induced by $\beta_3$, leading to a net reduction in charged-particle production.

The effect of Woods--Saxon parameters in extreme configurations can be further examined through comparison with the spherical case. Figure~5 shows the ratio of charged--hadron multiplicity in body--body and tip--tip collisions relative to the spherical configuration, for multiple parameter sets and two centrality classes. The main observations are summarized below.

In Fig.~5(a), for the 0--5\% centrality class, the case--2 ratio lies slightly above unity, whereas the case--1 ratio remains below. For Ru, which incorporates only quadrupole deformation, one might expect case--1, with its larger $\beta_{2}$, to produce a higher ratio in body--body collisions. However, the observed ordering indicates that the larger surface diffuseness in case--2 ($a = 0.485$~fm) results in a modestly increased overlap region and hence a higher $N_{\mathrm{ch}}$. This implies that, in central body--body collisions, surface diffuseness plays a more dominant role than quadrupole deformation. A similar trend is observed in the 50--60\% centrality bin, where the case--2 ratio again exceeds that of case--1. In peripheral collisions, particle production is primarily governed by the nuclear surface, and a larger diffuseness leads to a more extended overlap region and slightly enhanced charged--particle yields.

Figure~5(c) presents the tip--tip to spherical ratio for Ru+Ru collisions. In the 0--5\% centrality class, the ratios for case--1 and case--2 are nearly identical, with case--2 showing a marginally higher value. The reduced elongation combined with the larger surface diffuseness in case--2 leads to a slightly increased effective overlap in central tip--tip collisions. In contrast, in the peripheral (50--60\%) bin, the case--1 ratio exceeds that of case--2. In this regime, the smaller surface diffuseness in case--1 produces a comparatively denser overlap region, resulting in a modest enhancement of $N_{\mathrm{ch}}$ despite the larger diffuseness in case--2.

Figure~5(b) shows the body--body to spherical ratio for Zr+Zr collisions. Despite differences in $\beta_{2}$ and $a$, the ratios in the most central (0--5\%) collisions are nearly identical and close to unity, indicating that deformation effects are weak in this regime. This suggests that the inclusion of octupole deformation $\beta_{3}$ does not significantly modify charged--particle production in central body--body collisions. It should be noted that this reflects the limited sensitivity of this particular observable under consideration to deformation effects in the most central bin (0--5\%), rather than a general weakness of deformation effects, which are in fact expected to be strongest in central collisions and are more prominently reflected in other observables such as $v_2$. In peripheral collisions, the case--1 ratio increases, consistent with the reduced quadrupole deformation leading to a slightly larger effective overlap region. While a similar enhancement might be expected for case--2 based on geometric considerations alone, the comparatively lower ratio observed instead suggests that the shape asymmetry introduced by $\beta_{3}$ reduces particle production in this configuration.

In Fig.~5(d), the tip--tip to spherical ratios for Zr+Zr collisions are shown. In the 0--5\% centrality class, the ratios for case--1 and case--2 remain close, with case--2 exhibiting a slightly higher value, consistent with its larger surface diffuseness. However, in peripheral tip--tip collisions (50--60\%), the case--1 ratio becomes larger than that of case--2. Since the quadrupole deformation parameters in the two cases are similar and the surface diffuseness is larger in case--2, this suppression is non-trivial. The observed reduction of the case--2 ratio in peripheral tip--tip collisions is consistent with a negative contribution from the octupole deformation $\beta_{3}$, which introduces shape asymmetry and reduces the effective overlap region.

Taken together, the results presented in Figs.~3--5 demonstrate that nuclear deformation and surface diffuseness introduce systematic but modest modifications to charged--particle production across different collision configurations and centrality classes. For all configurations and centralities considered, the deformation--induced effects primarily manifest as variations in the overall charged--particle yield, while the pseudorapidity dependence remains largely unchanged. This behavior indicates that, within the present framework, the dominant influence of nuclear geometry arises from changes in the effective overlap region rather than from significant longitudinal modifications of the pseudorapidity distributions.

\subsection{$N_{\text{ch}}$ distribution analysis}

We have examined the centrality-dependent behavior of the $P(N_{\text{ch}})$ ratio between Ru and Zr for various collision configurations. $P(N_{\text{ch}})$ is the probability distribution of charged hadron multiplicity. 

In Fig. 6(b), the case-3 WS density parametrization is selected based on the recent calculations of energy density functional theory (DFT) \cite{Xu:2021vpn}, assuming the nuclei as spherical. The ratio $\frac{P(N_{\mathrm{ch}})^{\text{Ru}}}{P(N_{\mathrm{ch}})^{\text{Zr}}}$ rises at large multiplicities due to the extended high-$N_{\mathrm{ch}}$ tail in Ru+Ru collisions. This stems from the smaller root-mean-square size of Ru compared to Zr in DFT calculations, which leads to higher energy density and an increased number of binary nucleon–nucleon collisions in central events \cite{Li:2018oec}.  In mid-central to semi-peripheral regions ($N_{\text{ch}} = 50-300$), the ratios are consistent with the experimental trend. The dip in the ratio at peripheral collisions (between $N_{\text{ch}} = 0-50$) can be attributed to a thicker skin of Zr, which leads to a more diffuse distribution of peripheral density. 

A similar high-$N_{\text{ch}}$ trend, initially observed in the spherical configuration for $0$–$5$\% centrality, is also present in central tip–tip (case-1) [Fig. 6(c)] and body–body (case-2) [Fig. 7(a)] collisions. For peripheral ($50$–$60$\%) body–body collisions in case-2 [Fig. 7(a)], the ratio is well below unity, consistent with the earlier spherical-case explanation that a thicker nuclear surface yields a more diffuse peripheral density distribution. Meanwhile, for the tip-tip of case-1 [Fig. 6(c)], since the surface diffuseness is the same for both Ru and Zr, so the dip is not found in the peripheral collision zone. By analogous reasoning, the body--body collisions of case--1 [Fig.~6(a)] do not exhibit a pronounced suppression of the ratio in the peripheral (50--60\%) centrality range.

The body-body collisions of case-1 [Fig. 6(a)] and the tip-tip collisions of case-2 [Fig. 7(b)] do not exhibit the characteristic peak in higher multiplicity ($0$-$5$\%). We discuss each case individually below. In case--1, the surface diffuseness parameter is identical for Ru and Zr; therefore, the observed differences arise primarily from their respective deformation parameters. In central body--body collisions, Zr exhibits a slightly higher charged--hadron yield, consistent with its comparatively less elongated shape providing a larger transverse overlap region. In contrast, for tip--tip collisions in case--2, the surface diffuseness differs between the two systems. Ru has a smaller diffuseness parameter, whereas Zr possesses more diffuse nuclear tips. Consequently, central Zr+Zr collisions exhibit a modest enhancement in charged--hadron production relative to Ru+Ru. In mid-central to peripheral region ($N_{\text{ch}}$=50-300), the body-body collisions of case-1 [Fig. 6(a)] is below the experimental plot, while tip-tip (case-2) [Fig. 7(b)] is above it. This is most likely because, in body-body collisions, the Zr+Zr system forms a dense collision zone due to its less elliptic shape. While in the tip-tip case (case-2), due to the added asymmetry in Zr ($\beta_3$) the $\frac{P(N_{\mathrm{ch}})^{\text{Ru}}}{P(N_{\mathrm{ch}})^{\text{Zr}}}$ ratio stays above unity. 

The effect of Zr’s asymmetry is most evident in body–body collisions (case-2) [Fig. 7(a)], where both the most central and peripheral events can be explained by the same reasoning previously provided for spherical nuclei. Although the simulated body-body curve for case-1 [Fig. 6(a)] lies below unity in the mid-centrality range, the corresponding case-2 curve [Fig. 7(a)] exceeds unity in the same range. This indicates that the $\beta_{3}$ parameter in Zr reduces the production of charged hadrons in body-body collisions.

\subsection{$v_2$ ratio}

In Fig. 8, we have shown the centrality-wise elliptic flow ratio $\langle v_{2} \rangle(\mathrm{Ru})/\langle v_{2} \rangle(\mathrm{Zr})$ distribution for different WS parameters and different configurations. All curves corresponding to different parameter sets and configurations are compared with the STAR $v_{2}\{\mathrm{EP}\}$ data~\cite{STAR:2021mii}. 

In the body-body collision of case-1, The smaller quadrupole deformation in Zr leads to an enhancement of the $\langle v_{2} \rangle$ ratio (blue squares), hence the ratio remains above unity across centralities. While in the tip-tip collisions of case-1 (the green triangles), the ratio falls below unity. In tip-tip collisions, the nuclei overlap along the longer axis. A larger quadrupole deformation increases elongation along the major axis. When a collision happens between such elliptically deformed nuclei for a tip-tip configuration, it results in a more spherical fireball. Therefore, in case--1, the larger $\beta_{2}$ leads to a reduction in the final--state momentum anisotropy in the tip--tip configuration. This geometric effect is consistent with the $\langle v_{2} \rangle$ ratio remaining below unity across all centrality classes in tip--tip collisions of case--1. The pronounced dip observed in the tip-tip ratio around $40$-$50$\% centrality is a consequence of limited statistical sampling in that centrality bin rather than a genuine physical effect.

Next, we include the octupole deformation using case-2 of the parameter set. The tip–tip configuration without $\beta_{3}$ shows no substantial deviation from the corresponding distribution with $\beta_{3}$ (green triangles vs. pink rhombus). This suggests that the octupole deformation parameter does not produce a substantial modification of the elliptic flow in tip–tip collisions. However, the body–body curve for case-2 (red triangles) lies significantly below, falling even beneath the spherical case. This means the octupole deformation enhances $\langle v_2 \rangle$ in Zr+Zr collisions for the body-body configuration. Even in the spherical case, the ratio is less than one. Since both nuclei are completely round in this case, intrinsic nuclear deformation does not contribute to $\langle v_2 \rangle$. Thus, it can be concluded that surface diffuseness and centrality have a significant impact on the elliptic flow ratio in collisions of spherical nuclei.

\begin{figure}[h]
\includegraphics[width=.5\textwidth]{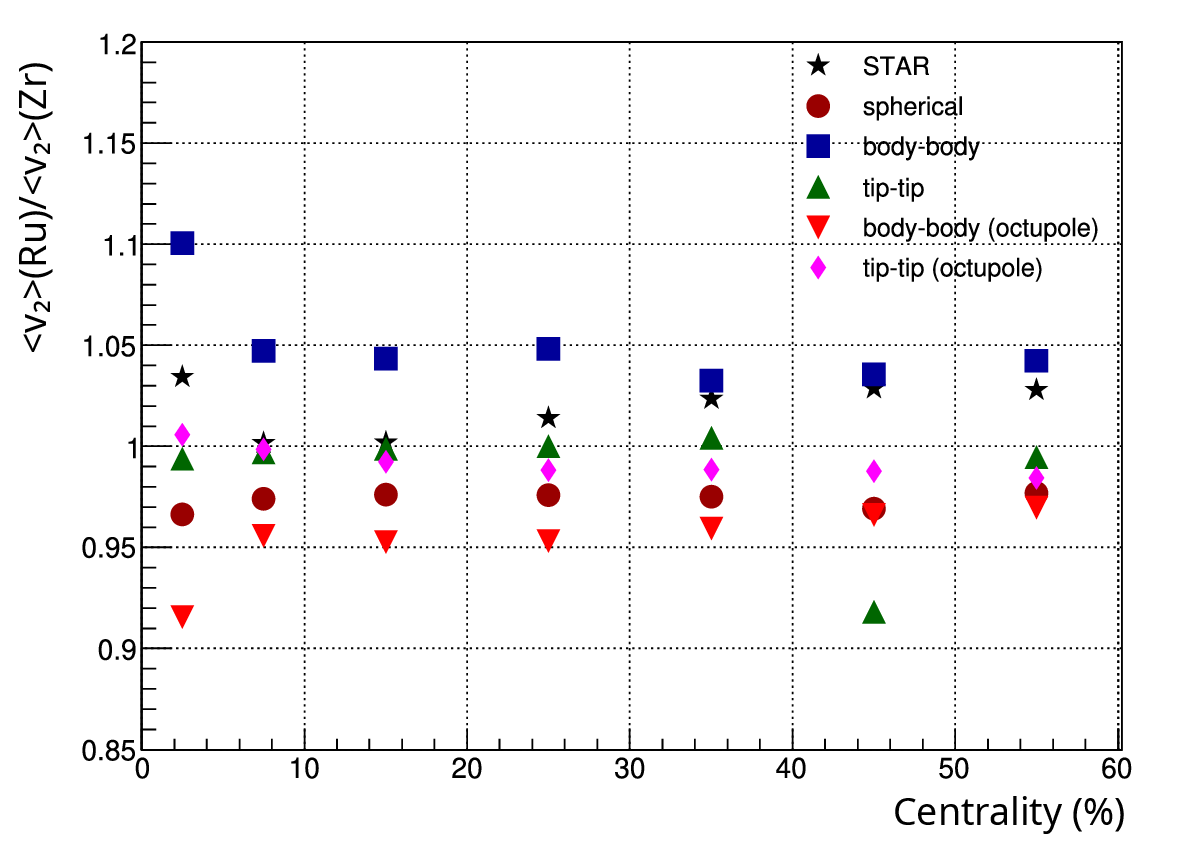}
\caption{\label{fig:2} Centrality-wise $v_2$ ratio distribution for different configurations. }
\end{figure}

\section{Summary}

The primary focus of most isobar collision studies concentrates on CME signatures like $\gamma$ correlator. However, nuclear deformation parameters, which significantly influence the flow-related background to CME signals, also need thorough investigation. In this work, we present a configuration--dependent analysis of Ru+Ru and Zr+Zr collisions within the HYDJET++ framework, systematically varying the quadrupole ($\beta_{2}$), octupole ($\beta_{3}$), and surface diffuseness ($a$) parameters across extreme geometric orientations. A key feature of our study is the determination of centrality separately for each collision configuration, enabling a controlled isolation of geometry-driven effects. By comparing deformed nuclei with their spherical counterparts, we disentangle the individual contributions of deformation and surface diffuseness to charged-particle production and elliptic flow. 

We found that the quadrupole deformation ($\beta_2$) and the surface diffuseness parameter ($a$) significantly affect the elliptic flow ($v_2$) and particle multiplicity. We have demonstrated that $\beta_3$ has a positive impact on the charged hadron production in the most-central ($0-5$\%) tip-tip collisions through analysis of the $ \eta$ distributions. Meanwhile, in peripheral tip-tip collisions ($50-60$\%), the effect of $\beta_3$ is negative. The ratio of $\eta$ distributions of deformed to spherical nuclei were also studied. In $50$-$60$\% centrality, $\beta_3$ has a negative effect on particle production for the body-body collisions. The effect of $\beta_3$ can also be seen on the $N_{\text{ch}}$ distribution with respect to centrality. The reduced charged hadron production in Zr suggests that $\beta_{3}$ negatively affects charged hadron production in mid-central ($N_{\text{ch}}$ = 50–300) body–body collisions. In centrality-wise $v_2$ distribution plots, the introduction of octupole deformation enhances the elliptic flow in Zr+Zr collisions. The effect of $\beta_3$ is more prominent in body-body collisions. A distinctive aspect of the present study is the configuration-dependent analysis of pseudorapidity distributions, which demonstrates that deformation effects primarily influence the overall particle yield while leaving the longitudinal shape largely unchanged.

Hence, the combined study of $\eta$ distributions, $N_{\mathrm{ch}}$ spectra, and centrality--dependent $\langle v_{2} \rangle$ ratio within a single framework provides a structured and internally consistent approach to quantifying deformation--induced background effects. This orientation--selected methodology offers a complementary perspective to conventional event--by--event analyses and may help refine the interpretation of CME--sensitive measurements in relativistic heavy--ion collisions. In realistic collisions, nuclei interact with random orientations. The observed multiplicity and elliptic flow represent an orientation-averaged result lying between the two extreme configurations. However, the sizable differences between these limits indicate that residual deformation-induced effects can remain non-negligible even after averaging. A proper account of these orientation-dependent geometric effects is therefore essential for a reliable interpretation of CME signals in isobaric collisions.

\section*{Data Availability}

Data and codes will be made available upon request.

\section*{Acknowledgements}

BKS sincerely acknowledges financial support from the
Institute of Eminence (IoE), BHU Grant number 6031. AD acknowledges the financial support through the
Institute fellowship from IIITDM Jabalpur. 
SRN acknowledges the financial support from the UGC
Non-NET fellowship and IoE research incentive during
the research work.

\end{document}